\documentclass[twocolumn, 
prb,preprintnumbers,amsmath,amssymb,floatfix]{revtex4}

\usepackage{graphicx}
\usepackage{dcolumn}

\begin{document}

\title{Antiphase magnetic boundaries in iron-based superconductors: A first-principle density-functional theory study}
\author{Z. P. Yin and W. E. Pickett}
\affiliation{Department of Physics, University of California Davis, 
  Davis, CA 95616}
\date{\today}
\begin{abstract}
Superconductivity arises in the layered iron-pnictide compounds  
when magnetic long range order disappears.  
We use first principles density functional methods to 
study magnetic arrangements that may compete with long range order near the phase
boundary.  Specifically, we study the energetics and charge density distribution
(through calculation of the electric field gradients) for ordered supercells with
varying densities of antiphase magnetic boundaries.  We quantify the amount
by which Fe atoms with 
low spin moments at the antiphase boundaries have higher energies than Fe atoms with
high spin moments away from the antiphase boundaries.
These disruptions in magnetic order should be useful in accounting for 
experimental data such as 
electric field gradients and hyperfine fields on both Fe and As atoms. 
\end{abstract}
\maketitle

\section{Background and Motivation}
The discoveries of high temperature superconductivity in the FeAs-based layered compounds
\cite{Kamihara-LaO, Chen-CeO,Ren-NdO,Ren-PrO, Cheng-GdO, Wang-GdO, Ren-SmO, Rotter-Ba, Sasmal-Sr,Torikachvili-Ca,
Tapp-Li, Parker-Na, Matsuishi-SrF}
of doped or compressed ZrCuSiAs-type $R$FeAsO and $X$FeAsF, ThCr$_2$Si$_2$-type $M$Fe$_2$As$_2$, 
and Cu$_2$Sb-type $A$FeAs
($R$: rare-earth metal; $X$=Ca, Sr; $M$=Ca, Sr, Ba, Eu; $A$=Li, Na),
with critical temperatures up to 56 K, has been attracting excitement in the condensed matter community.
In the vicinity of room temperature,
these compounds crystallize in tetragonal symmetry with no magnetic order. 
At some lower temperature (which can be in the range of 100 - 210 K),
they undergo a first or second order phase transition to an orthorhombic structure and become antiferromagnetically ordered. 
\cite{Cruz-La,McGuire-La-FeEFG,Kitagawa-Ba-AsEFG, Baek-Ca-AsEFG}
The structural transition and magnetic order transition can happen simultaneously or successively depending on the compound.
\cite{Cruz-La,Kitagawa-Ba-AsEFG, Baek-Ca-AsEFG}
It was confirmed both experimentally and theoretically that the magnetic order of Fe at low temperature
 is stripe-like antiferromagnetism often referred to as spin density wave (SDW).
\cite{Cruz-La,Kitagawa-Ba-AsEFG, Baek-Ca-AsEFG, Yin-La, Yildirim-La}
Upon doping or compressing, the magnetic order goes away and the materials become superconducting. 

Much theoretical work has been reported since the first discovery
\cite{Kamihara-LaO} of LaFeAsO$_{1-x}$F$_x$, 
with many aspects of these compounds having been 
addressed,\cite{Yin-La, Yildirim-La, Singh-La, seb-njp, Mazin-La, Ma-La, GGA-magnetism, Haule-La}
but with many questions unresolved.  
A central question is what occurs at the
SDW-to-SC (superconducting) phase transition, and what drives this change, and more fundamentally what microscopic
pictures is most useful in this enterprise.  In the $R$FeAsO compounds, doping with carriers
of either sign leads to this transition, even though there seems little that is special about the
band-filling in the stoichiometric compounds.  In the $M$Fe$_2$As$_2$ system, the SDW-to-SC transition
can be driven with pressure (relatively modest, by research standards) without any doping whatever,
apparently confirming that doping level is not an essential control parameter.
Some delicate characteristic seems to be involved, and one way of addressing the loss of magnetic
order is to consider alternative types of magnetic order, and their energies.

Many results, experimental and theoretical, indicate itinerant magnetism in this system, and LSDA calculations
without strong interaction effects included correctly predict the type of antiferromagnetism observed.
There is however the general feature that 
the calculated ordered moment of Fe is larger than the observed value.
For example, neutron scattering experiment\cite{Cruz-La} obtained the ordered Fe magnetic moment of
0.36 $\mu$$_B$ in LaFeAsO, while calculations\cite{Yin-La,seb-njp,GGA-magnetism} result in 
the much larger values 1.8-2.1 $\mu_B$.
Neutron diffraction and neutron scattering experiments
{\cite{Su-Ba, Zhao-Sr, Goldman-Ca} estimated the Fe magnetic moment in the SDW state 
of $M$Fe$_2$As$_2$ ($M$=Ba, Sr, Ca)
to be in the range of 0.8 $\mu$$_B$ to 1.0 $\mu$$_B$ but our calculations (this work) give  
1.6-1.9 $\mu_B$.
This kind of (large) discrepancy of the ordered magnetic moment is unusual in Fe-based magnets,
and there are efforts underway to understand the discrepancy as well as the mechanism underlying
magnetic intereactions.\cite{johannes09}
In addition, $^{57}$Fe M\"{o}ssbauer experiments
\cite{McGuire-La-FeEFG,Kitao-La-FeEFG, Klauss-La-FeEFG, Nowik-La-FeEFG, Nowik-Re-FeEFG, Ba-FeEFG, Sr-FeEFG,
SrF-FeEFG} and $^{75}$As NMR measurements
\cite{Kitagawa-Ba-AsEFG, Baek-Ca-AsEFG,EFG-LaAs} further  
confirm the disagreement in magnetic moments and electric field gradients 
between experiments and ab initio calculations in the SDW state.

To explain these significant disagreements, it is likely  
that spin fluctuations in some guise play a role in these compounds.  
The SDW instability is a common interpretation of the magnetic order in these compounds,
\cite{Chen-CeO, Cruz-La,Drew-Sm, Hu-Ba} 
which implicates 
the influence of spin fluctuation in the magnetically disordered state.  
An inelastic neutron scattering study on a single-crystal sample of BaFe$_2$As$_2$ by Matan {\it et al.}
\cite{Matan-Ba} showed anisotropic scattering around the antiferromagnetic wave vectors,
suggestive of two dimensional spin fluctuation in BaFe$_2$As$_2$.  Such possibilities must 
be reconciled with the existence of
high energy spin excitations in the SDW state of BaFe$_2$As$_2$ as observed by Ewings  {\it et al.}
\cite{Ewings-Ba}.

One of the simplest spin excitations is that arising from antiphase boundaries in the SDW phase. 
Mazin and Johannes have introduced such ``antiphasons" and their dynamic fluctuations as being
central for understanding the various phenomena observed in this class of materials.\cite{Mazin-NP}
The structural transition followed by the antiferromagnetic transition, 
the change of slope and a peak in the differential resistivity $d\rho (T)/dT$ at the phase transitions,
and the invariance of the resistivity
anisotropy over the entire temperature range can be qualitatively understood in their scenario by considering
dynamic antiphase boundaries (twinning of magnetic domains).\cite{Mazin-NP}
\begin{figure}[tbp]
{\resizebox{7.8cm}{5.6cm}{\includegraphics{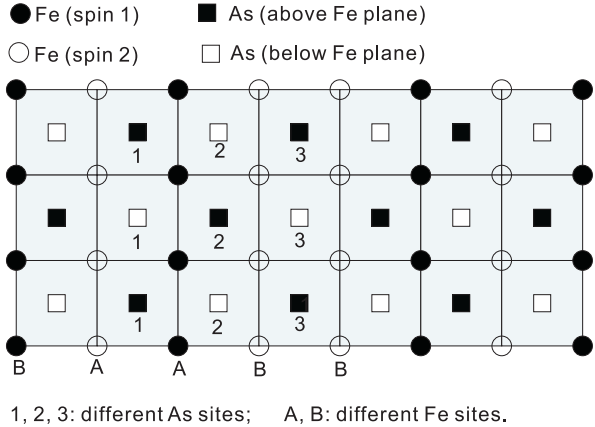}}}
\caption{The structure of FeAs layer in the Q-SDW state showing the antiphase boundary in the
center of the figure. Fe spin 1 (filled circle) and spin 2 (empty circle) 
have two different sites A (`bulklike') and B (`boundary-like'). As above Fe plane 
(filled square) and below Fe plane (empty square) have three sites 1, 2, 3 whose local
environments differ.}
\label{QQMstructure}
\end{figure}

In this paper, we consider a class of magnetic arrangements derived from the stripe-like AFM phase:
static periodic magnetic arrangements (SDWs) representing antiphase boundaries that require 
doubled, quadrupled, and octupled supercells.
We denote these orders as D-SDW, Q-SDW and O-SDW, respectively. Figure \ref{QQMstructure} shows
the magnetic arrangements of Fe in the Q-SDW phase.
Its unit cell is a 4$\times$1$\times$1 supercell of the SDW unit cell. 
Antiphase boundaries occur at the edge and the center of its unit cell along a-axis (antiparallel/alternating Fe spins), the same as in
the D-SDW and O-SDW states, 
the unit cells of which are 2$\times$1$\times$1 and 8$\times$1$\times$1 supercells of the SDW unit cell, respectively.
The D-SDW phase can also  be viewed as a double stripe SDW phase.
Based on the results of 
these states, we consider the effect of
antiphase boundary spin fluctuations in explaining various experimental results, 
which was discussed to some extent by Mazin and Johannes.\cite{Mazin-NP}
The antiphase magnetic boundaries we consider here are the simplest possible and yet  
explain semiquantitatively many experimental results by assuming that the dynamic average over antiphase magnetic 
boundaries can be modeled by averaging over several model antiphase boundaries.
Our picture presumes that the antiphase boundary within the magnetically ordered state  
is in some sense a representative spin  
excitation in the FeAs-based compounds.

The calculations are done using the linear augmented plane wave (LAPW) method as implemented in the Wien2k code \cite{wien2k}, 
with both PW91\cite{PW91} and PBE\cite{PBE} exchange-correlaton (XC) potentials.  
Several results have been double checked using the full potential local orbital (FPLO) i
code\cite{fplo1, fplo2} 
with PW91 XC potential.

\section{LDA VS. GGA}
Whereas the local density approximation (LDA) for the exchange-correlation (XC)
potential usually obtains internal coordinates accurately, 
it has been found\cite{seb-njp, GGA-magnetism} 
that LDA makes unusually large errors when predicting the 
As height $z$(As) in these compounds in either nonmagnetic (NM) or SDW states.  The generalized gradient approximation (GGA)
makes similar errors in the NM state, 
however, GGA predicts very good values of $z$(As) in the SDW phase, as shown in 
Fig. \ref{zAs-errors}. 
One drawback of GGA is that it enhances magnetism\cite{GGA-magnetism,seb-njp} 
in these compounds over the LDA prediction, which is already too large compared to its observed value. 
For example, using experimental structural parameters, 
GGA (PBE) gives a Fe spin magnetic moment larger than LDA (PW91) by 0.3 $\mu$$_B$ in the SDW state,
 and more than 0.6  $\mu$$_B$ in the D-SDW state. 
The magnetic moment changes the charge density, roughly in proportion to the moment.
For this reason, we have used Wien2K with PW91 (LDA) XC functional 
with its more reasonable moments to calculate
the EFG and hyperfine field using experimental structural parameters. 
We note that  PBE and PW91 produce about the same EFG in the NM state.
\begin{figure}[tbp]
\rotatebox{-90}
{\resizebox{6.0cm}{7.8cm}{\includegraphics{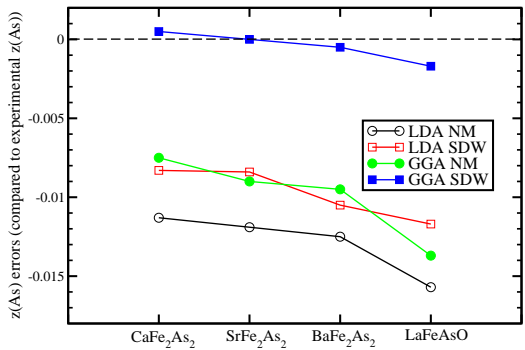}}}
\caption{The calculated errors of z(As) compared to experimental values in the NM and SDW states when using LDA (PW91) and GGA (PBE)
XC functionals in CaFe$_2$As$_2$,  SrFe$_2$As$_2$,  BaFe$_2$As$_2$, and LaFeAsO.}
\label{zAs-errors}
\end{figure}

\section{The magnetic moment and hyperfine field of iron}
In the various antiphase boundary SDW states, the Fe atoms can assume two different characters: high-spin A site
away from the antiphase boundaries and low-spin B site at the antiphase boundaries, as shown in Fig. \ref{QQMstructure}. 
The Fe atoms with the same site (A or B) in these states have about the same magnetic moment and hyperfine field.
For example, in BaFe$_2$As$_2$ in the static
Q-SDW state, the spin magnetic moment and hyperfine field for Fe at A site are 1.59 $\mu$$_B$ and 12.6 T,
and for Fe at B site are 0.83 $\mu$$_B$  and 6.2 T.
In the O-SDW state, the spin magnetic moments and hyperfine fields for the three (slightly different)
A sites are 1.59, 1.60, 1.67 $\mu$$_B$ and 12.6, 13.5, 1.37 T, respectively, 
and are 0.83 $\mu$$_B$ and 6.1 T for Fe at B site.

As mentioned above, there are significant differences of the ordered magnetic moment of Fe in the SDW state between
calculated values and values observed in neutron scattering (diffraction) experiments and/or M\"{o}ssbauer experiments.
Table \ref{hyperfinefield} shows Fe spin magnetic moment and hyperfine field in the SDW and static D-SDW state calculated
by FPLO7 and Wien2K using PW91 XC functional. 
\begin{table}
\caption{The experimental magnetic moment of Fe m$_{Fe}$ (in unit of $\mu$$_B$)
 and the hyperfine field B$_{hf}$ (in unit of Tesla) for Fe,
and values calculated in the SDW and D-SDW ordered phases,
using Wien2K with PW91 for the $M$Fe$_2$As$_2$ ($M$=Ba, Sr, Ca), LaFeAsO and SrFeAsF compounds.
The experimental values are in all cases much closer to the D-SDW values (with its maximally dense antiphase boundaries)
than to the SDW values.
\cite{Cruz-La,Klauss-La-FeEFG, Goldman-Ca,Su-Ba, Zhao-Sr, Ba-FeEFG, Sr-FeEFG, SrF-FeEFG}
For the Fe magnetic moment, results from both FPLO (denoted as FP) and Wien2k(denoted as WK) are given.
Because these methods (and other methods) differ somewhat in their assignment of the moment to an Fe
atom, the difference gives some indication of how strictly a value should be presumed.
}
\label{hyperfinefield}
\begin{tabular}{|c|c|c|c|c|c|c|c|c|}
\hline
      & \multicolumn{2}{|c|}{exp} & \multicolumn{3}{|c|}{SDW} & \multicolumn{3}{|c|}{D-SDW}\\
\hline
compound & m$_{Fe}$ & B$_{hf}$ & \multicolumn{2}{|c|}{m$_{Fe}$} & B$_{hf}$ & \multicolumn{2}{|c|} {m$_{Fe}$} & B$_{hf}$ \\
         &          &          & FP & WK &WK & FP & WK & WK\\
\hline
BaFe$_2$As$_2$ & 0.8  &5.5 &1.78 & 1.65 &13.6 & 0.90 & 0.80 & 5.4 \\
SrFe$_2$As$_2$ & 0.94 &8.9 &1.80 & 1.68 &13.9 & 0.98 & 0.91 &6.1  \\
CaFe$_2$As$_2$ & 0.8  & -- &1.63 & 1.53 &12.4 & 0.77 & 0.71 & 4.7  \\
LaFeAsO        & 0.36  & -- &1.87 & 1.77 &14.9 & 0.50 & 0.48 & 3.6 \\
SrFeAsF        & ---  & 4.8& --  &1.66 &14.5  & ---  & 0.32 &2.3  \\
\hline
\end{tabular}
\end{table}

\section{Energy differences}
Since Fe atoms in all these antiphase boundary SDW states have basically two spin states 
(high-spin state at A sites and low-spin state at B sites),
in a local moment picture one might expect that energy differences could be related to just
the two corresponding energies.  From our comparison of energies we have found
that this picture gives a useful 
account of the energetic differences. 

The SDW and the D-SDW state are the simple cells in this regard. The former 
doesn't have any low spin Fe 
and the latter doesn't have any high spin Fe, so these two define the high
spin (low energy) and low spin (high energy) ``states'' of the Fe atom.
Table \ref{SDWenergy} shows the total energies per Fe (in meV) 
of the magnetic phases compared to the 
non-magnetic stater, for BaFe$_2$As$_2$,
SrFe$_2$As$_2$, CaFe$_2$As$_2$, LaFeAsO and SrFeAsF. 
The high spin and low spin energies  
vary from  system to system. The energy of the Q-SDW state has also been
calculated, and it can be compared with the average of high spin and low
spin moment energies (last column in Table \ref{SDWenergy}).  The reasonable
agreement indicates that corrections beyond this simple picture are minor.

The energy cost to create an antiphase boundary is (roughly)
simply the cost of two
extra low spin Fe atoms versus the high spin that would result without the
antiphase boundary.  This difference is found to vary
by over a factor of two, in the range of 40-90 
meV per Fe for this set of five compounds. The reason for the variation is not
apparent; for example, it is not directly proportional to the Fe moment (or its square). 

\begin{table}
\caption{Calculated  total energies (meV/Fe) compared to NM state of the 
various SDW states (SDW, D-SDW, Q-SDW) in 
the $M$Fe$_2$As$_2$ ($M$=Ba, Sr, Ca), LaFeAsO and SrFeAsF compounds. 
The energy tabulated in the last column, labeled Q$^{\prime}$, is the 
average of the high spin (SDW) and low spin (D-SDW) energies.
illustrating that the energy of the Q-SDW ordered phase
follows this average reasonably
well. The level of agreement indicates to what degree `high spin' 
and `low spin' is a reasonable
picture of the energetics at an antiphase boundary.
}
\label{SDWenergy}
\begin{tabular}{|c|c|c|c|r|}
\hline
compound       & SDW   & D & Q & Q$^{\prime}$  \\
\hline
BaFe$_2$As$_2$ & -73 & -6  & -36 & -39    \\
SrFe$_2$As$_2$ & -91 & -11 & -46 & -51    \\
CaFe$_2$As$_2$ & -66 & -8  & -33 & -37    \\
LaFeAsO        & -143 & -61 &-94 & -102   \\
SrFeAsF        & -73 & 0  & -40  & -37    \\
\hline
\end{tabular}
\end{table}
Analogous calculations were also carried out for the large O-SDW cell for BaFe$_2$As$_2$.
As for the other compounds and antiphase supercells, the Fe moments could be characterized
by a low spin atom at the boundary and high spin Fe elsewhere.  The energy could also be
accounted for similarly, analogously to Table \ref{SDWenergy}.

\section{The Electric Field Gradient}
The EFG at the nucleus of an atom is sensitive to the anisotropy of the electron charge distribution
around the atom. A magnitude and/or symmetry change of the EFG implies the local environment around the atom changes,
which can be caused by changes in bonding, structure, or magnetic ordering. In BaFe$_2$As$_2$ during the simultaneously 
structure transition from tetragonal to orthorhombic and magnetic order transition from non-magnetic to SDW order at about 135 K,
the EFG component V$_c$ along the crystal c-axis drops rapidly by 10$\%$ 
and the asymmetry parameter
\begin{math}
\eta=\frac{|V_{a}-V_{b}|}{|V_{c}|}
\end{math}
jumps from zero to larger than one, indicating the principle axis for the largest component V$_{zz}$ is changed from along c-axis to
in the ab plane.\cite{Kitagawa-Ba-AsEFG}
The abrupt EFG change reflects a large change in the electron charge distribution around As sites and
highly anisotropic charge distribution in the ab plane. 
A similar thing happens in CaFe$_2$As$_2$ except that the V$_c$ component of
CaFe$_2$As$_2$ in the non-magnetic state is {\it five times} that in BaFe$_2$As$_2$, and 
doubles its value at the structural and magnetic
transition at 167 K when it goes to the SDW phase.\cite{Baek-Ca-AsEFG}
The different behaviour of the EFG change across the phase transition in BaFe$_2$As$_2$ and CaFe$_2$As$_2$ may be due to
the out-of-plane alkaline-earth atom (Ba and Ca in this case), which influences the charge distribution around As atoms. 
It also indicates that three dimensionality is more important in $M$Fe$_2$As$_2$ than in $R$FeAsO, 
which is evident in the layer distance of the FeAs layers reflected in the c lattice constant
of these compounds. The c lattice constant of CaFe$_2$As$_2$ is significantly smaller than BaFe$_2$As$_2$, 
so that the interlayer interaction of the FeAs layers is stronger in CaFe$_2$As$_2$, therefore the charge distribution
in CaFe$_2$As$_2$ is more three dimensional like than in BaFe$_2$As$_2$, which can be clearly seen in their Fermi surfaces (not shown). 
 
\subsection{EFG of As atoms}
The EFG of As can be obtained from the quadrupole 
frequency in nuclear quadrupolar resonance (NQR) measurement 
in nuclear magnetic resonance (NMR) experiment.
The NQR frequency can be written as
\begin{equation}
{\nu_Q=\frac{3eQV_{zz}(1+\eta^2/3)^{1/2}}{2I(2I-1)h}}, 
\end{equation}
where Q ($\sim$ 0.3b, b=10$^{-24}$ cm$^2$) is the $^{75}$As quadrupolar moment, V$_{zz}$ is the zz component of As EFG,
\begin{equation}
\eta=\frac{|V_{xx}-V_{yy}|}{|V_{zz}|}
\end{equation}
 is the asymmetry parameter of the EFG, I=3/2 is the $^{75}$As nuclear spin, $h$ is the Planck constant. In LaO$_{0.9}$F$_{0.1}$FeAs,
Grafe {\it et al.}\cite{EFG-LaAs} reported $\nu$$_Q$=10.9 MHz, and $\eta$=0.1, which gives V$_{zz}$ $\sim$ 3.00 $\times$ 10$^{21}$ V/m$^2$.
(Note: for the value of EFG, the unit 10$^{21}$ V/m$^2$ is commonly used, and we will adopt this unit for all EFG values below)
The experimental value 3.0 agrees satisfactorily with our result\cite{seb-njp} of 2.7 calculated by Wien2K code. 

In BaFe$_2$As$_2$ and CaFe$_2$As$_2$,  NMR experiments suggest
the V$_c$ component of the EFGs of As are 0.83 and 3.39 respectively at high temperature 
in the NM states,\cite{Kitagawa-Ba-AsEFG, Baek-Ca-AsEFG} 
while our calculated values are 1.02 and 2.35, respectively. 
The difference is in the right direction and right order of magnitude though not quantitatively accurate.
However, the EFGs calculated in the SDW state don't match  experimental observations at all. 
In BaFe$_2$As$_2$ from 135 K down to very low temperature, V$_c$ remains around 0.62 and
\begin{math}
\eta=\frac{|V_{a}-V_{b}|}{|V_{c}|}
\end{math}
changes from 0.9 to 1.2. Our calculated results in the SDW state gives 
V$_a$=1.34, V$_b$=-1.47, and V$_c$=-0.13,
which gives $\eta$ $\approx$ 20.
The calculated results substantially 
underestimate V$_c$ and overemphasize the anisotropy in the ab plane.

We now consider whether 
these discrepancies can be clarified if antiphase boundary is considered. 
In the static D-SDW, Q-SDW and O-SDW states, 
the surrounding environment of
As sites change. 
Depending on the magnitudes (high spin or low spin) and directions (parallel or antiparallel) 
of the spins of their nearest and next nearest neighboring Fe atoms, As atoms generally have
three different sites (1, 2, and 3) as shown in Fig. \ref{QQMstructure}.
In the static D-SDW state,
As atoms have similar site 1$^{\prime}$ and 3$^{\prime}$. 
As shown in Table \ref{BaAs-EFG}, 
the calculated quantities for these states cannot directly explain the experimental observed values neither.

However, they may be understandable
if the antiphase boundary is dynamic, i.e., any As atom in a given measurement can change from site 1 to site 2 and/or site 3,  
when its nearby Fe atoms flip their spin directions. These time fluctuations have be represented by
an average over configurations, and we consider briefly what arises from a configuration average
over our SDW phase and three short-period ordered cells (having different densities of 
anti-phase boundaries).
In the O-SDW state, for example,
if an As samples 25$\%$, 25$\%$ and 50$\%$ ``time'' as As sites 1, 2 and 3, respectively 
during the experimental measurement, 
then the expectation values are V$_a$=0.99, V$_b$=-0.29, V$_c$=-0.71, 
$\eta$=1.80, B$_{hf}$=1.43 T. This simple consideration
already match much better with experimental\cite{Kitagawa-Ba-AsEFG} observed 
V$_c$ $\sim$ 0.62, H$_{in}$=1.4 T,
 except $\eta$ which is around 1.2.
The actual situation could be much more complicated, being an average over all 
the sites in all the static
D-SDW, Q-SDW and O-SDW states, and other more complicated states. 
Considering the
relatively small differences of the EFGs at the same site for Q-SDW and O-SDW order,
As sites 
in other static antiphase
boundary SDW states should be able to be classified to site 1, 2 and 3 as in 
the Q-SDW and O-SDW states.

\begin{table}
\caption{The calculated EFG component V$_a$, V$_b$, V$_c$ (in unit of 10$^{21}$ V/m$^2$) , 
the asymmetry parameter $\eta$, spin magnetic moment of As ($\mu$$_B$), hyperfine field at the As nuclei (Tesla) 
of BaFe$_2$As$_2$ in the SDW, D-SDW, Q-SDW and O-SDW states. Experimentally, V$_c$ is around 0.62, $\eta$ is in the range
of 0.9 to 1.2, and the internal field at As site parallel to c axis is about 1.4 T.\cite{Kitagawa-Ba-AsEFG} 
See text for notation.}
\label{BaAs-EFG}
\begin{tabular}{|c|c|c|c|c|c|c|}
\hline
state &site & V$_a$ & V$_b$& V$_c$ & $\eta$  &B$_{hf}$ \\
\hline
SDW   &1 & 1.21 &-1.32 &0.11 &23.0      &0   \\
\hline
D-SDW &1$^{\prime}$ &0.63 &0.07 &-0.70 &0.8     & 0  \\
      &3$^{\prime}$ & 0.66 &0.37 &-1.03 &0.28  &2.1 \\
\hline
Q-SDW &1 &1.17 &-1.38 &0.21 &12.1  &0 \\
      &2 &0.99 &-0.72 &-0.27 &6.33 & 1.0 \\
      &3 &0.90 &0.51 &-1.41 &0.28  &2.1 \\
\hline
O-SDW &1 &1.25 & -1.44 &0.19 &14.2  &0 \\
      &2 &0.98 & -0.71 &-0.27 &6.26 &1.1 \\
      &3 &0.87  & 0.50 &-1.37 &0.27 &2.3 \\
\hline
\end{tabular}
\end{table}

\subsection{EFG of Fe atoms}
The EFG of Fe can be obtained from the electric quadrupole splitting parameter derived from 
M\"{o}ssbauer measurements. The electric quadrupole splitting parameter can be written as 
\begin{equation}
{\Delta=\frac{3eQV_{zz}(1+\eta^2/3)^{1/2}}{2I(2I-1)}}, 
\end{equation}
which equals E$_{\gamma}$$\Delta$E$_Q$/c, where Q $\sim$ 0.16b is the $^{57}$Fe 
quadrupolar moment,  E$_{\gamma}$ is the energy
of the $\gamma$ ray emitted by the $^{57}$Co/Rh source, $\Delta$E$_Q$ is the electric 
quadrupole splitting parameter from
M\"{o}ssbauer data given in the unit of speed, and c is the speed of light in vacuum. 
By fitting the M\"{o}ssbauer spectra, one also obtains the isomer shift $\delta$ and average
hyperfine field B$_{hf}$.

We consider whether the dynamic antiphase boundary spin fluctuation picture can also 
clarify the comparison between calculated and observed EFG of Fe.
We take SrFe$_2$As$_2$ as an example. 
As shown in Table \ref{SrFe-EFG}, 
in the NM state, the calculated value V$_Q$=0.98 agrees well with the V$_Q \approx$ 0.83 at room temperature.
V$_Q$ calculated in the SDW state (0.68) agrees rather well with experimental value about 0.58 at 4.2 K.
V$_Q$ calculated in the D-SDW state (0.80) is somewhat larger than that in the SDW state, 
but not .
Regarding EFG, the biggest difference between SDW and D-SDW is the asymmetry 
parameter--it is 0.61 in the SDW and only 0.11 in the latter.
Further experiments are required to clarify this difference.

\begin{table}
\caption{The calculated EFG component V$_a$, V$_b$, V$_c$ (in unit of 10$^{21}$ V/m$^2$) ,
the asymmetry parameter $\eta$=$|$V$_{xx}$-V$_{yy}$$|$/$|$V$_{zz}$$|$ 
(here $|$V$_{zz}$$|$ $>$ $|$V$_{xx}$$|$ and $|$V$_{yy}$$|$), V$_Q$=$|$V$_{zz}$$|$/(1+$\eta$$^2$/3)$^{1/2}$
of Fe in SrFe$_2$As$_2$ in the NM, SDW, D-SDW, Q-SDW and O-SDW states. 
Experimentally, V$_Q$ is around 0.83 at room temperature
in the non-magnetic state, and it is about 0.58 at 4.2 K.\cite{Sr-FeEFG} }
\label{SrFe-EFG}
\begin{tabular}{|c|c|c|c|c|c|c|}
\hline
state &site & V$_a$ & V$_b$& V$_c$ & $\eta$ & V$_Q$ \\
\hline
NM    &0 &-0.49 &-0.49 & 0.98 & 0 & 0.98 \\
\hline
SDW   & A&-0.58 & -0.14 & 0.72 & 0.61 & 0.68 \\
\hline
D-SDW &B & -0.30 & -0.51 & 0.81 & 0.26 & 0.80 \\
\hline
Q-SDW &A &-0.63 &-0.06 &0.69 &0.83 & 0.62 \\
      &B &-0.56 &-0.49 &1.05 &0.07 & 1.05\\
\hline
\end{tabular}
\end{table}

\section{Summary}
Experiments generally indicate that an itinerant magnetic moment, magnetic (SDW) instability, 
and spin fluctuations are common features of the Fe-based superconductors. 
In this paper, we have studied the energetics, charge density distribution
(through calculation of the electric field gradients, hyperfine fields and magnetic moments) for ordered supercells with
varying densities of antiphase magnetic boundaries, namely the SDW, D-SDW, Q-SDW and 
(for very limitied cases) O-SDW phases.
Supposing dynamic magnetic antiphase boundaries are present, and that the 
spectroscopic experiments average over them, 
we can begin to clarify several seemingly contradictory experimental and computational results.  

Our calculations tend to support the idea that antiphase boundary magnetic configurations can be
important in understanding data.  The fact that the decrease in moment is confined to the
antiphase boundary Fe atom does not mean that a local moment picture is appropriate; in fact,
exactly this same type of local spin density calculations provide a description of magnetic
interaction that is at odds with a local moment picture.\cite{johannes09}
The calculated energy cost to create an antiphase
boundary is however rather high for the cases we have considered, 
and this value would seem to restrict formation of antiphase boundaries at temperatures
of interest. Calculations that treat actual disorder, and dynamics as well, would be very
helpful in furthering understanding in this area.

\section{Acknowledgments}
This work was supported by DOE grant DE-FG02-04ER46111.  We also acknowledge support from
the France Berkeley Fund that enabled the initiation of this project.


\begin{thebibliography}{10}

\bibitem{Kamihara-LaO} Y. Kamihara,
  T. Watanabe, M. Hirano, and H. Hosono,
   J. Am. Chem. Soc. {\bf 130}, 3296 (2008).

\bibitem{Chen-CeO}G. F. Chen, Z. Li, 
 D. Wu, G. Li, W. Z. Hu, J. Dong, P. Zheng, J. L. Luo, and N. L. Wang, 
Phys. Rev. Lett. {\bf 100}, 247002 (2008).

\bibitem{Ren-NdO}Z.-A. Ren, J. Yang, W. Lu, 
 W. Yi, X.- L. Shen, Z.-C. Li, G.-C. Che, X.-L. Dong, L.-L. Sun, F. Zhou, and Z.-X. Zhao, 
Europhysics Letters {\bf82} (2008) 57002 

\bibitem{Ren-PrO}Z.-A. Ren, J. Yang, W. Lu,
 W. Yi, G.-C. Che, X.-L. Dong, L.-L. Sun, and Z.-X. Zhao, 
Materials Research Innovations {\bf 12}, 105, (2008)

\bibitem{Cheng-GdO}P. Cheng, L. Fang, H. Yang,
  X. Y. Zhu, G. Mu, H. Q. Luo, Z. S. Wang, and H.-H. Wen, 
Science in China G 51(6), 719-722(2008).

\bibitem{Wang-GdO}C. Wang, L. J. Li, S. Chi, 
 Z. W. Zhu, Z. Ren, Y. K. Li, Y. T. Wang, X. Lin, Y. K. Luo, S. Jiang, X. F. Xu, 
 G. H. Cao, and Z. A. Xu, 
Europhysics Letters {\bf 83}, 67006 (2008).

\bibitem{Ren-SmO}Z.-A. Ren, W. Lu, J. Yang, 
 W. Yi, X.-L. Shen, Z.-C. Li, G.-C. Che, X.-L. Dong, L.-L. Sun, F. Zhou, and Z.-X. Zhao, 
Chin. Phys. Lett. {\bf 25}, 2215 (2008).

\bibitem{Rotter-Ba}M. Rotter, M. Tegel, and D. Johrendt, Phys. Rev. Lett. {\bf 101}, 107006 (2008).

\bibitem{Sasmal-Sr}K. Sasmal, B. Lv, B. Lorenz, A. M. Guloy, F. Chen, Y. Y. Xue, and C. W. Chu, Phys. Rev. Lett. {\bf 101}, 107007 (2008)

\bibitem{Torikachvili-Ca}M. S. Torikachvili, S. L. Bud'ko, N. Ni, and P. C. Canfield, Phys. Rev. Lett. {\bf 101}, 057006 (2008).

\bibitem{Tapp-Li}J. H. Tapp, 
 Z. J. Tang, B. Lv, K. Sasmal, B. Lorenz, P. C.W. Chu, and A. M. Guloy, 
Phys. Rev. B {\bf 78}, 060505(R) (2008).

\bibitem{Parker-Na}D. R. Parker, M. J. Pitcher, P. J. Baker, I. Franke, T. Lancaster, S. J. Blundell, and S. J. Clarke, 
Chem. Commun. (Cambridge) {\bf 2009}, 2189.

\bibitem{Matsuishi-SrF}S. Matsuishi, Y. Inoue, T. Nomura, M. Hirano, and H. Hosono,
J. Phys. Soc. Jpn. {\bf 77}, 113709 (2008).

\bibitem{Cruz-La} C. de la Cruz, Q. Huang, 
 J. W. Lynn, J. Li, W. R. II, J. L. Zarestky, H. A. Mook, G. F. Chen, J. L. Luo, and N. L. Wang,
  Nature {\bf 453}, 899  (2008).

\bibitem{McGuire-La-FeEFG}M. A. McGuire, 
 A. D. Christianson, A. S. Sefat, B. C. Sales, M. D. Lumsden, R. Y. Jin, 
 E. A. Payzant, D. Mandrus, Y. B. Luan, V. Keppens, V. Varadarajan, J. W. Brill, 
 R. P. Hermann, M. T. Sougrati, F. Grandjean, and G. J. Long, 
Phys. Rev. B {\bf 78}, 094517 (2008).

\bibitem{Kitagawa-Ba-AsEFG} K. Kitagawa, N. Katayama, K. Ohgushi, M. Yoshida, and M. Takigawa, 
J. Phys. Soc. Jpn. {\bf 77}, 114709 (2008).


\bibitem{Baek-Ca-AsEFG}S.-H. Baek, N. J. Curro, T. Klimczuk, E. D. Bauer, F. Ronning, and J. D. Thompson, 
Phys. Rev. B {\bf 79}, 052504 (2009).

\bibitem{Yin-La} Z. P. Yin,
     S. Leb\'{e}gue, M. J. Han, B. P. Neal, S. Y. Savrasov, and W. E. Pickett,  
  Phys. Rev. Lett. {\bf 101}, 047001 (2008).

\bibitem{Yildirim-La} T. Yildirim,
	Phys. Rev. Lett. {\bf 101}, 057010 (2008).

\bibitem{Singh-La} D. J. Singh and M.-H. Du,
   Phys. Rev. Lett. {\bf 100}, 237003 (2008).

\bibitem{seb-njp}S. Leb\'{e}gue, Z. P. Yin, and W. E. Pickett, 
New J. Phys. {\bf 11}, 025004 (2009).

\bibitem{Mazin-La}I. I. Mazin,
    D. J. Singh, M. D. Johannes, and M. H. Du,
Phys. Rev. Lett. {\bf 101}, 057003 (2008).

\bibitem{Ma-La} F. Ma, Z.-Y. Lu, and T. Xiang, 
Phys. Rev. B {\bf 78}, 224517 (2008).

\bibitem{GGA-magnetism}I. I. Mazin,
    M. D. Johannes, L. Boeri, K. Koepernik, and D. J. Singh,
   Phys. Rev. B {\bf 78}, 085104 (2008).

\bibitem{Haule-La}K. Haule, J. H. Shim, and G. Kotliar, Phys. Rev. Lett. {\bf 100}, 226402 (2008).


\bibitem{Su-Ba}Y. Su,
 P. Link, A. Schneidewind, T. Wolf, P. Adelmann, Y. Xiao, M. Meven, R. Mittal, M. Rotter, D. Johrendt, 
 T. Brueckel, and M. Loewenhaupt, 
Phys. Rev. B {\bf 79}, 064504 (2009).

\bibitem{Zhao-Sr}J. Zhao,
 W. Ratcliff, J. W. Lynn, G. F. Chen, J. L. Luo, N. L. Wang, J. Hu, and P. Dai, 
Phys. Rev. B {\bf 78}, 140504(R) (2008).


\bibitem{Goldman-Ca} A.I. Goldman,
 D.N. Argyriou, B. Ouladdiaf, T. Chatterji, A. Kreyssig, S. Nandi, N. Ni, S. L. Bud'ko, P.C. Canfield, and R. J. McQueeney, 
Phys. Rev. B {\bf 78}, 100506(R) (2008).

\bibitem{johannes09}M. D. Johannes and I. I. Mazin, Phys. Rev. B {\bf 79},
220510(R) (2009).

\bibitem{Kitao-La-FeEFG}S. Kitao, 
 Y. Kobayashi, S. Higashitaniguchi, M. Saito, Y. Kamihara, M. Hirano, T. Mitsui, H. Hosono, and M. Seto, 
J. Phys. Soc. Jpn. {\bf 77} 103706 (2008)

\bibitem{Klauss-La-FeEFG}H.-H. Klauss, 
 H. Luetkens, R. Klingeler, C. Hess, F.J. Litterst, M. Kraken, M. M. Korshunov, 
 I. Eremin, S.-L. Drechsler, R. Khasanov, A. Amato, J. Hamann-Borrero, 
 N. Leps, A. Kondrat, G. Behr, J. Werner, and B. Buchner, 
Phys. Rev. Lett. {\bf 101}, 077005 (2008).

\bibitem{Nowik-La-FeEFG}I. Nowik, I. Felner, V. P. S. Awana, A. Vajpayee, and H. Kishan, 
J. Phys.: Condens. Matter {\bf 20}, 292201 (2008).

\bibitem{Nowik-Re-FeEFG}I. Nowik and I. Felner, J. Supercond. Nov. Magn. {\bf 21}, 297 (2008). 


\bibitem{Ba-FeEFG}M. Rotter, M. Tegel and D. Johrendt, 
I. Schellenberg, W. Hermes, and R. P\"{o}ttgen,
Phys. Rev. B {\bf 78}, 020503(R) (2008).

\bibitem{Sr-FeEFG}M. Tegel, 
M. Rotter, V. Weiss, F. M. Schappacher, R. Poettgen, and D. Johrendt, 
J. Phys.: Condens. Matter {\bf 20}, 452201 (2008).

\bibitem{SrF-FeEFG}M. Tegel, 
 S. Johansson, V. Weiss, I. Schellenberg, W. Hermes, R. Poettgen, and D. Johrendt
Europhys. Lett. {\bf 84}, 67007 (2008).

\bibitem{EFG-LaAs} H.-J. Grafe,
 D. Paar, G. Lang, N. J. Curro, G. Behr, J. Werner, J. Hamann-Borrero, C. Hess, N. Leps, R. Klingeler, and B. B$\ddot{u}$chner,
Phys. Rev. Lett.  {\bf 101}, 047003 (2008).

\bibitem{Drew-Sm}A. J. Drew, Ch. Niedermayer, P. J. Baker, F. L. Pratt, S. J. Blundell, T. Lancaster, 
R. H. Liu, G. Wu, X. H. Chen, I. Watanabe, V. K. Malik, A. Dubroka, M. R\"{o}essle, 
K. W. Kim, C. Baines, and C. Bernhard, 
Nature Mater. {\bf 8}, 310 (2009).

\bibitem{Hu-Ba}W. Z. Hu, J. Dong, 
G. Li, Z. Li, P. Zheng, G. F. Chen, J. L. Luo, and N. L. Wang, 
Phys. Rev. Lett. {\bf 101}, 257005 (2008).

\bibitem{Matan-Ba}K. Matan, R. Morinaga, K. Iida, and T. J. Sato, 
Phys. Rev. B {\bf 79}, 054526 (2009).  


\bibitem{Ewings-Ba}R.A. Ewings, 
T.G. Perring, R.I. Bewley, T. Guidi, M.J. Pitcher, D.R. Parker, S.J. Clarke, and A.T. Boothroyd, 
Phys. Rev. B {\bf 78}, 220501(R) (2008).

\bibitem{Mazin-NP}I. I. Mazin and M. D. Johannes, Nature Physics {\bf 5}, 141-145 (2009).

\bibitem{wien2k}P. Blaha,
  K. Schwarz, G. K. H. Madsen, D. Kvasnicka, and J. Luitz,
   \textsf{WIEN2K},(K. Schwarz, Techn. Univ. Wien, Austria) (2001).

\bibitem{PW91} J. P. Perdew and Y. Wang,
   Phys. Rev. B {\bf 45}, 13244 (1992).

\bibitem{PBE} J. P. Perdew, K. Burke, and M. Ernzerhof,
   Phys. Rev. Lett {\bf 77}, 3865 (1996).

\bibitem{fplo1}K. Koepernik and H. Eschrig,
   Phys. Rev. B {\bf 59}, 1743 (1999).

\bibitem{fplo2}K. Koepernik,
  B. Velicky, R. Hayn, and H. Eschrig,
   Phys. Rev. B {\bf 55}, 5717 (1997).

\end{thebibliography}
\end{document}